\documentclass[final,smallextended]{svjour3}     

\usepackage{graphicx}
\usepackage{mathtools}
\usepackage{booktabs}
\usepackage{subcaption}
\usepackage{newtxtext,newtxmath}
\usepackage{natbib}
\usepackage{multirow}
\usepackage{xcolor}
\usepackage{siunitx}
\usepackage{bm}
\usepackage{hyperref}
\usepackage[left=2cm, right=2cm, bottom=2cm, top=2.5cm]{geometry}
\newcommand{\z}[1]{\textcolor{black}{#1}}

\DeclarePairedDelimiter{\norm}{\lVert}{\rVert}

\def\makeheadbox{{%
		\hbox to0pt{\vbox{\baselineskip=10dd\hrule\hbox
				to\hsize{\vrule\kern3pt\vbox{\kern3pt
						\hbox{\bfseries\textcolor{red}{Published in ``Experimental Astronomy''}}
						\hbox{\textcolor{red}{DOI: 10.1007/s10686-020-09687-4}}
						\kern3pt}\hfil\kern3pt\vrule}\hrule}%
			\hss}}}

\begin{document}

\title{Experimental evaluation of complete safe coordination of astrobots for Sloan Digital Sky Survey V}

\titlerunning{Experimental evaluation of complete coordination of astrobots for SDSS-V}

\author{Matin Macktoobian$^{1,\ast}$
	\and
	Ricardo Araújo$^{2}$
	\and 
	Loïc Grossen$^{2}$
	\and 
	Luzius Kronig$^{1}$
	\and 
	Mohamed Bouri$^{1}$
	\and 
	Denis Gillet$^{1}$
	\and 
	Jean-Paul Kneib$^{2,3}$
}

\authorrunning{Matin Macktoobian et al.}

\institute{$^{1}$ School of Engineering, Swiss Federal Institute of Technology in Lausanne (EPFL), 1015 Lausanne, Switzerland\\
           \and
           $^{2}$School of Basic Sciences, Swiss Federal Institute of Technology in Lausanne (EPFL), 1015 Lausanne, Switzerland\\
           \and 
           $^{3}$Observatory of Sauverny, 1290 Versoix, Switzerland\\$^{*}$\email{matin.macktoobian@epfl.ch}
}

\date{Received: Oct. 1st, 2020 / Accepted: Nov. 10th, 2020 / Published: Dec. 19th, 2020}

\maketitle

\begin{abstract}
The data throughput of massive spectroscopic surveys in the course of each observation is directly coordinated with the number of optical fibers which reach their target. In this paper, we evaluate the safety and the performance of the astrobots coordination in SDSS-V by conducting various experimental and simulated tests. We illustrate that our strategy provides a complete coordination condition which depends on the operational characteristics of astrobots, their configurations, and their targets. Namely, a coordination method based on the notion of cooperative artificial potential fields is used to generate safe and complete trajectories for astrobots. Optimal target assignment further improves the performance of the used algorithm in terms of faster convergences and less oscillatory movements. Both random targets and galaxy catalog targets are employed to observe the coordination success of the algorithm in various target distributions. The proposed method is capable of handling all potential collisions in the course of coordination. Once the completeness condition is fulfilled according to initial configuration of astrobots and their targets, the algorithm reaches full convergence of astrobots. Should one assign targets to astrobots using efficient strategies, convergence time as well as the number of oscillations decrease in the course of coordination. Rare incomplete scenarios are simply resolved by trivial modifications of astrobots swarms' parameters.
\end{abstract}

\section{Introduction}
\subsection{Massive spectroscopic surveys and SDSS-V}
Dark energy studies \citep{joyce2016dark} have been revolutionized once the accelerated expansion of the universe was observed \citep{riess1998observational}. In particular, the evolution of the universe, which has been under intense scrutiny over the recent decades, is found to be correlated with the distribution of dark matter all over the cosmos. A data-driven strategy to obtain the desired distribution requires abundant mass-energy recording of the universe. Accordingly, the map of the observable universe is expected to convey valuable information about the geometry and the evolution of the cosmos. \z{Universe expands over time}. Thus, redshift-based observation strategies effectively provide various volumes of the space in the course of different cosmological era associated with the age of the universe. In particular, the measurements of baryonic acoustic oscillations (BAO) \citep{seo2003probing} have already shed light on the filament-void interactions of the cosmos \citep{forero2009dynamical}. So, BAO analysis with respect to various redshift ranges is known to generate spectroscopic surveys which eventually yield a significant repertoire of the cosmological data to investigate the universe's evolution.
\begin{figure}[!htp]
	\centering
	\begin{subfigure}{.4\textwidth}
		\centering
		\includegraphics[scale=0.2]{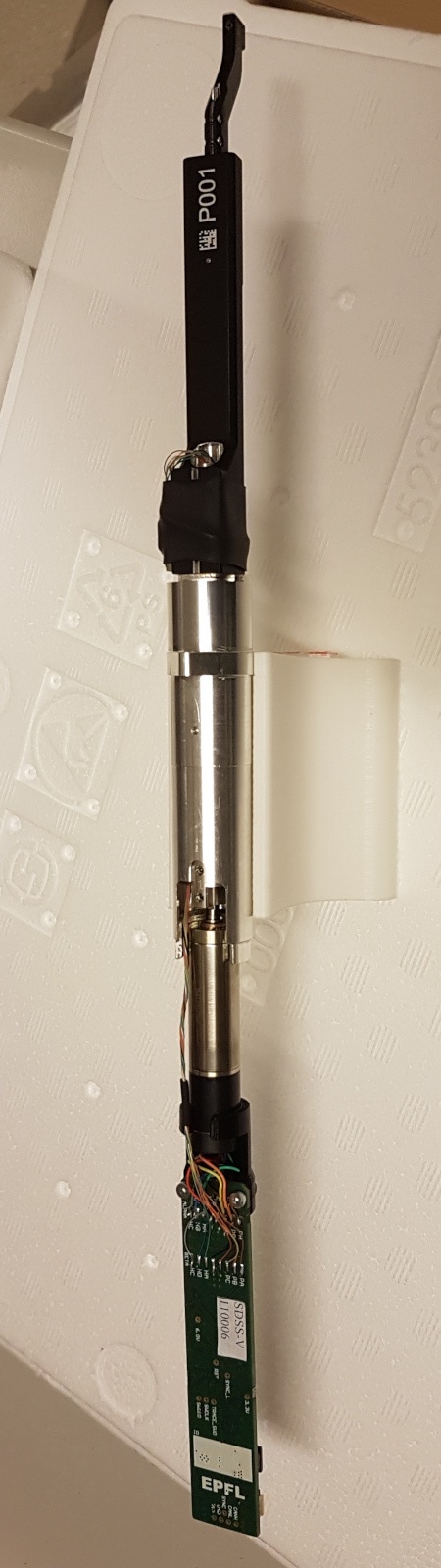}
		\caption{The astrobot side view\label{fig:side}}%
	\end{subfigure}
	\begin{subfigure}{.4\textwidth}
		\centering
		\subcaptionbox{The astrobot top view\label{fig:top}}
		{\includegraphics[scale=0.5]{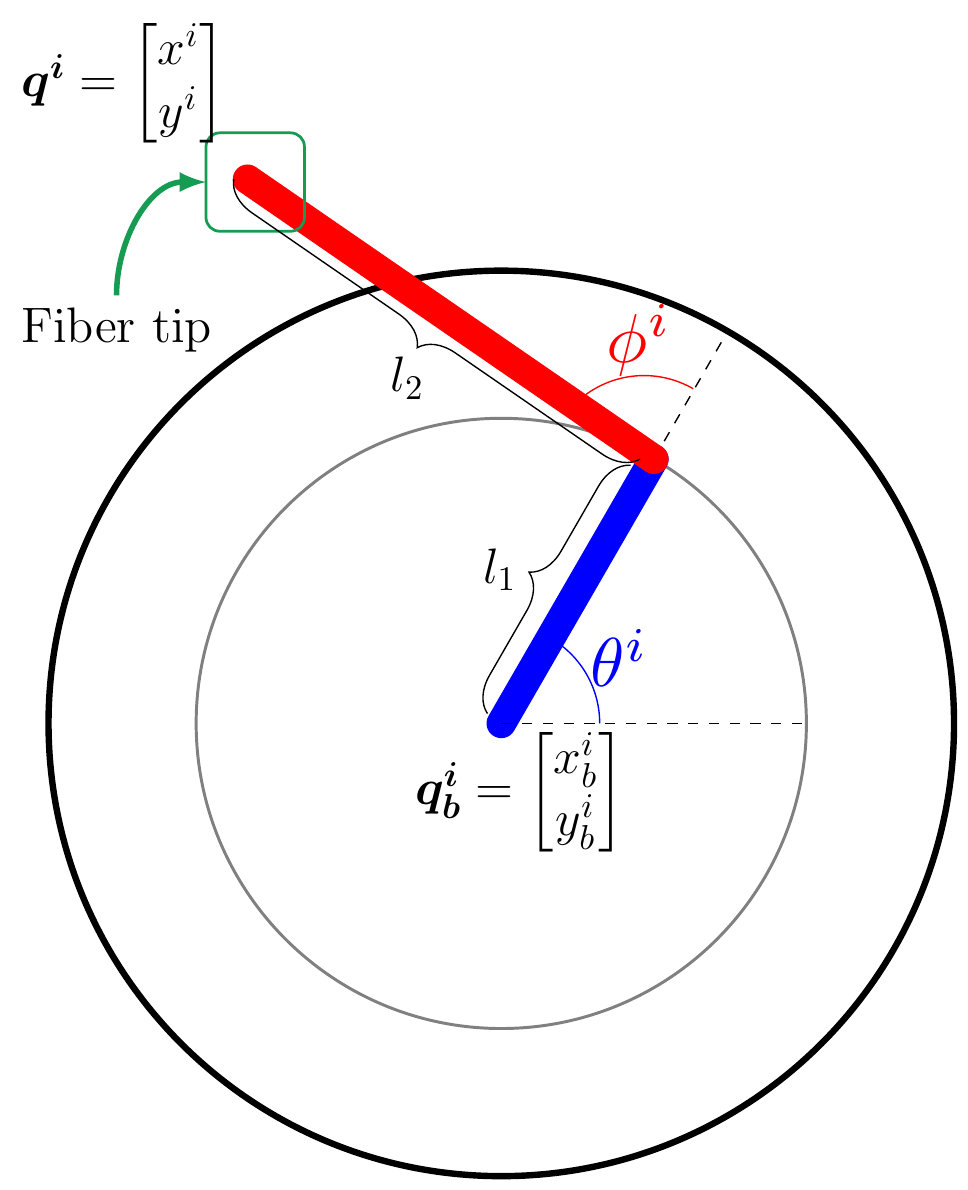}}
		\subcaptionbox{The focal plane schematic\label{fig:foc}}
		{\includegraphics[scale=0.08]{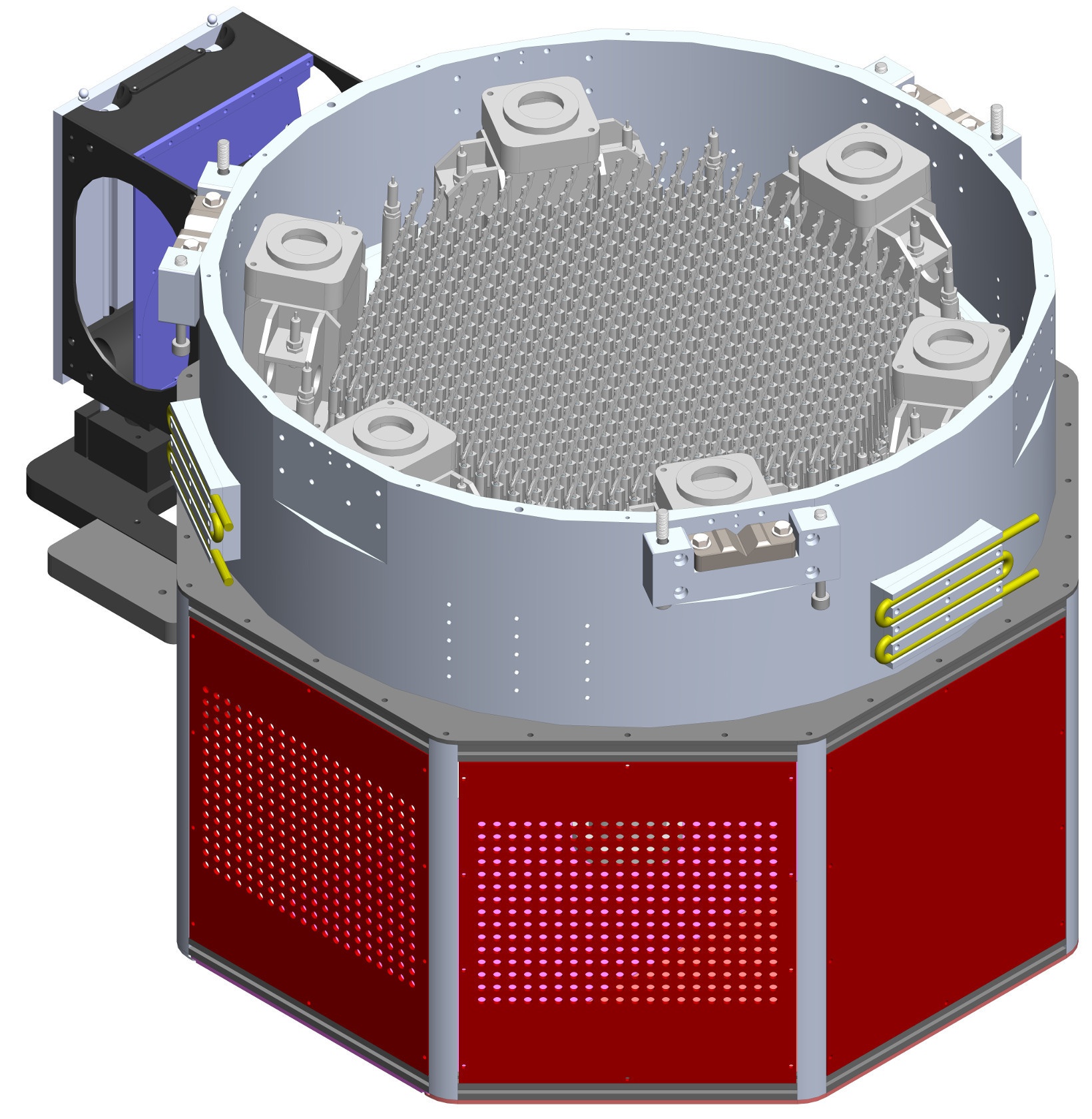}}
	\end{subfigure}
	\caption{A typical SDSS-V astrobot and overall focal plane (figures (b) and (c) are reprinted from \citet{macktoobian2019complete} and the SDSS-V wiki, respectively.)}
\end{figure}

\z{Each spectroscopic survey is defined based on a particular redshift range in which either matter or dark energy mostly dominates}. Namely, low-redshift surveys \citep{hamuy2006carnegie} extensively contributes to the study of dark energy, while high-redshift ones \citep{takada2006cosmology} are often the target to investigate the universe when its radiation is dominated by its mass. To be specific, the Anglo-Australian Telescope (AAT) \citep{blake2011wigglez} was used to generate surveys associated with star-forming galaxies at lower redshifts, say, $z < 0.8$. Middle-range redshifts, such as $0.7<z<2.2$, have been widely taken into account by various instruments, e.g., SDSS telescope. The resulting baryonic oscillation spectroscopic survey \citep{dawson2012baryon} is supposed to cover almost 1.4 million galaxies and quasars. High-redshift spectroscopic surveys (at $2<z<5$) are of utmost importance in the space-time inflation studies and dark energy observations \citep{schlegel2019megamapper}, as studied by \citet{ferraro2019inflation}. The combination of BAO analysis and redshift-space distortions \citep{scoccimarro2004redshift} seems to be promising enough to evaluate General Relativity with respect to various cosmological scales \citep{schlegel2019astro2020}. To this aim, massive spectroscopic surveys have to be planned which essentially include huge numbers of optical fibers, multi-segment focal planes, and vast telescope apertures.  

Sloan Digital Sky Survey (SDSS) represents a family of spectroscopic projects aiming to the generation of surveys using various observational technologies in different ground telescopes. Early 2000s witnessed the first generation of these projects, i.e., SDSS-I \citep{zehavi2002galaxy}. This project, based at APO, covered various spectral bands using camera-based photometry. APO later hosted SDSS-II \citep{ivezic2008milky} which yielded the first class of the multi-object optical spectroscopic surveys of the SDSS family. This project exclusively used optical fibers to collect visible lights of its desired targets. Then, the researchers' attention was shifted to near-infrared rays of targets which can also be captured by ground telescopes. Consequently, SDSS-III \citep{alam2015eleventh} ended up with the APOGEE survey supported by a near-infrared multi-object spectrograph at APO. The first three classes of SDSS projects were executed at APO in the northern hemisphere. To cover data acquisition from the southern hemisphere, SDSS-IV \citep{bundy2014overview} collected spectroscopic data, based on a regime similar to its predecessors, at Las Campanas Observatory (LCO). This project extensively covered many surveys such as APOGEE-2 \citep{zasowski2017target}, MaNGA \citep{wake2017sdss}, and the extended Baryon oscillation spectroscopic surveys \citep{dawson2016sdss}. SDSS-V \citep{kollmeier2017sdss} is about to conduct multi-object spectroscopic surveys in both hemispheres, namely, using Sloan Foundation telescope at APO in the northern hemisphere \citep{gunn20062} and the du Pont telescope at LCO in the southern hemisphere \citep{way2005redshifts}. Both types of optical and near-infrared fibers are used in this project which will be fed into a pair of optical BOSS and near-infrared APOGEE spectrographs. The resulting surveys are expected to extensively contribute to the characterization of Milky Way galaxy.

Massive spectroscopic survey projects such as DESI \citep{dey2019overview}, MOONS \citep{cirasuolo2014moons}, PFS \citep{takada2014extragalactic}, LAMOST \citep{cui2012large}, and in particular SDSS-V, include hundreds to thousands of fibers to maximize the information throughout of each observation. For this purpose, one may increase the aperture size of a host telescope to have a larger focal plane, thereby covering larger number of fibers. Second, fibers have to be placed in more dense formations to tile the area of the focal plane with higher observational resolution. However, fiber multiplexing raises operational challenges. In particular, fibers have to point to different locations of their fields from one observation to another. So, their reconfiguration needs to be performed in the available spare time between consecutive observations. To hit this mark, fibers were manually replaced in early versions of the surveys using SDSS spectrograph \citep{smee2013multi}. However, given the gradual increment of the employed fibers, robot fiber placement was taken into account in the case of AAT spectrograph. The cited process was inefficient because of the lack of any parallelism in the coordination of fibers. \z{Then, the first generation of robotic multi-fiber spectrographs showed up in the MX spectrometer \citep{hill1986deployment}, while recent instruments, such as fiber multi-object spectrograph (FMOS) \citep{kimura2010fibre}, LAMOST \citep{zhao2012lamost}, and MOONS \citep{cirasuolo2016moons}, are equipped with more fibers whose coordination are more challenging}. 
\subsection{Astrobots and safe complete coordination problem}
The state-of-the-art \z{astrobot} design \citep{horler201624mm}, called astrobot (see, Fig.\ref{fig:side} and \ref{fig:top}), is used in SDSS-V to automatically coordinate astrobots according to their target assignment plan. This \z{astrobot} is based on a $\theta-\phi$ design structure. In particular, each astrobot is a SCARA-like \citep{das2005mathematical} two-degree-of-freedom double-joint manipulator through which a fiber is passed. The fiber tip is located as the end-effector area of its astrobot, called ferrule. The motion of the ferrule in the patrol zone of the astrobot provides the fiber accessibility to various locations of the patrol zone which may host any observational target. The position of a ferrule can be formally modeled according to the rotational kinematics of the arms associated with it with respect to the base of its astrobot as follows.
\begin{equation}\label{eq:kin}
\bm{q^{i}} = \bm{q^{i}_{b}} + 
\begin{bmatrix}\cos{\theta^{i}} & \cos{(\theta^{i} + \phi^{i})}\\
\sin{\theta^{i}} & \sin{(\theta^{i} + \phi^{i})}\end{bmatrix} \bm{l}
\end{equation}
Here, vectors $\bm{q^{i}}$ and $\bm{q^{i}_{b}}$ are the coordinates of the ferrule and the base of $i$th astrobot, respectively. Vector $\bm{l}$ includes the lengths of the first and the second arms. 
To maximize the focal plane coverage, astrobots are located in hexagonal dense formation next to each other. On the other hand, one needs to have access any point in the focal plane using astrobots. Thus, the overall lengths of the two arms of each astrobot, known as pitch, has to be able to reach the center of any of its neighboring peers. Interested readers in a detailed treatment of the hardware characteristics of astrobots may refer to \citet{horler201624mm,macktoobian2019navigation}.This requirement guarantees the nominal reachability of any target in the scope of focal plane, but it also brings serious possibilities of deadlocks and collisions among neighboring astrobots. The coordination performances of astrobots are critical similarly to their safety. Astrobots have only a limited amount of time to be coordinated before an observation is initiated. In the case of massive focal planes, any coordination solution has to be fast enough not to jeopardize the on-time preparation of any desired configuration of astrobots. 

The first solution to the coordination of astrobots fulfills safety using the notion of navigation functions \citep{makarem2014collision} for \z{DESI} astrobots \citep{mathews2006method}. Nonetheless, the convergence rate of astrobots were not satisfactory. Namely, this method could barely coordinate less than 80\% of astrobots. Low convergence rates yield low-resolution surveys. Thus, \citet{tao2018priority} revised the cited strategy by adding a new automaton-based decision layer to the navigator of the algorithm. This layer directly resolves the deadlocks, which can not be handled by the main navigation function, based on a priority-based set of criteria. Thus, convergence rate reached $\sim$85\%. In the course of the design phase of SDSS-V, \citet{macktoobian2019complete,macktoobian2019navigation} proposed a new navigation function, i.e., cooperative artificial potential field (CAPF), to establish a completeness condition for a typical coordination based on the positional characteristics of astrobots and their targets. Thanks to this formulation, the global completeness convergence analysis of astrobots has shown to be localized in neighborhoods. In other words, one only has to check the completeness condition in all neighborhoods, instead of simultaneously solving hundreds of differential equations corresponding to the whole pack of astrobots.

In this paper, we evaluate the experimental behavior of the completeness seeker algorithm (CSA) in the scope of SDSS-V requirements. We also simulate and coordinate larger focal planes to have a glimpse of the CSA efficiency to be applied to future giant spectroscopic surveys, such as MegaMapper \citep{schlegel2019megamapper} which will be composed of $\sim$20,000 astrobots. Section \ref{subsec:theo} briefly outlines the control algorithm at the kernel of which CAPF is used. We describe the test bench of this study in Section \ref{subsec:setup}. The results of the applied experiments are explained in Section \ref{sec:res}. We draw our conclusion in Section \ref{sec:conc}.
\section{Methods\protect\footnote{Throughout this paper, scalars are represented by regular symbols. Bold symbols are reserved to denote matrices.}}
\subsection{Theory}
\label{subsec:theo}
Complete coordination of astrobots intend to achieve the maximum data \z{throughput} of an observation. A traditional navigation function includes an attractive and a repulsive term \citep{makarem2014collision}. The attractive term of an assigned navigation function to an astrobot attracts it to its target, whereas the repulsive one avoids collisions between the astrobot and its surrounding counterparts. Accordingly, an astrobot's velocity profile is proportional to the gradient of its potential. Then, once the astrobot reaches its target, the gradient vanishes, thereby stopping at its goal point. Navigation functions of this strategy are selfish, in that once a robot ends up at its target, it refrains any further movements. However, it may block the paths of its peers to reach their targets. In this scenario, the convergence rate may be diverged from achieving the total coordination of astrobots. 

Alternatively, the complete coordination theory of astrobots \citep{macktoobian2019complete} takes a less-selfish scheme into account in that each astrobot is not completely indifferent to the convergence success of its neighbors. According to this formalism, the global convergence of a swarm of astrobots directly depends on the local convergences corresponding to all of their neighborhoods. In particular, each neighborhood encompasses an astrobot in addition to all of its neighboring peers. For this purpose, a CAPF $\xi(\bm{q^{i}})$, say, Eq. (\ref{eq:CAPF}), is assigned to the $i$th astrobot which includes not only attractive and repulsive terms but also a cooperative attractive term varying according to convergence statuses of its neighbors. 
\begin{equation}
\label{eq:CAPF}
%\begin{split}
%\xi(\bm{q^{i}}) := &\underbrace{\vphantom{\bm{\lambda_{2}}\displaystyle\sum\limits_{j \in \mathcal{I}_{\mathcal{N}^{i}}\setminus \{i\}} \min [0,\frac{\norm{\bm{q^{i}} - \bm{q^{j}}}^{2} - D^{2}}{\norm{\bm{q^{i}} - \bm{q^{j}}}^{2} - d^{2}}}\bm{\lambda_{1}}\norm{\bm{q^{i}} - \bm{q^{i}_{\mathcal{T}}}}^{2}}_{\text{attractive term}} + \underbrace{\bm{\lambda_{3}}\displaystyle\sum\limits_{\mathclap{j \in \mathcal{I}_{\mathcal{N}^{i}}\setminus \{i\}}}\norm{\bm{q^{j}} - \bm{q^{j}_{\mathcal{T}}}}^{2}}_{\text{cooperative attractive term}} + \\& \underbrace{\bm{\lambda_{2}}\displaystyle\sum\limits_{\mathclap{j \in \mathcal{I}_{\mathcal{N}^{i}}\setminus \{i\}}} \min~ \left[\bm{0},\frac{\norm{\bm{q^{i}} - \bm{q^{j}}}^{2} - D^{2}}{\norm{\bm{q^{i}} - \bm{q^{j}}}^{2} - d^{2}}\right]}_{\text{repulsive term}}
%\end{split}
\xi(\bm{q^{i}}) := \underbrace{\vphantom{\bm{\lambda_{2}}\displaystyle\sum\limits_{j \in \mathcal{I}_{\mathcal{N}^{i}}\setminus \{i\}} \min [0,\frac{\norm{\bm{q^{i}} - \bm{q^{j}}}^{2} - D^{2}}{\norm{\bm{q^{i}} - \bm{q^{j}}}^{2} - d^{2}}}\bm{\lambda_{1}}\norm{\bm{q^{i}} - \bm{q^{i}_{\mathcal{T}}}}^{2}}_{\text{attractive term}} + \underbrace{\bm{\lambda_{3}}\displaystyle\sum\limits_{\mathclap{j \in \mathcal{I}_{\mathcal{N}^{i}}\setminus \{i\}}}\norm{\bm{q^{j}} - \bm{q^{j}_{\mathcal{T}}}}^{2}}_{\text{cooperative attractive term}} + \underbrace{\bm{\lambda_{2}}\displaystyle\sum\limits_{\mathclap{j \in \mathcal{I}_{\mathcal{N}^{i}}\setminus \{i\}}} \min~ \left[\bm{0},\frac{\norm{\bm{q^{i}} - \bm{q^{j}}}^{2} - D^{2}}{\norm{\bm{q^{i}} - \bm{q^{j}}}^{2} - d^{2}}\right]}_{\text{repulsive term}}
\end{equation}   	
Here, $\bm{q^{i}}$ and $\bm{q^{i}_{\mathcal{T}}}$ are the current and the target positions of the $i$th astrobot; $\lambda_{1}$, $\lambda_{2}$, and $\lambda_{3}$ are weighting factors; $\mathcal{N}^{i}$ is the neighborhood of the $i$th astrobot; $D$ denotes the collision avoidance envelope radius of each astrobot; $d$ represents the safety radius around each astrobot.
\begin{figure}[!htp]
	\centering
	\includegraphics[scale=0.85]{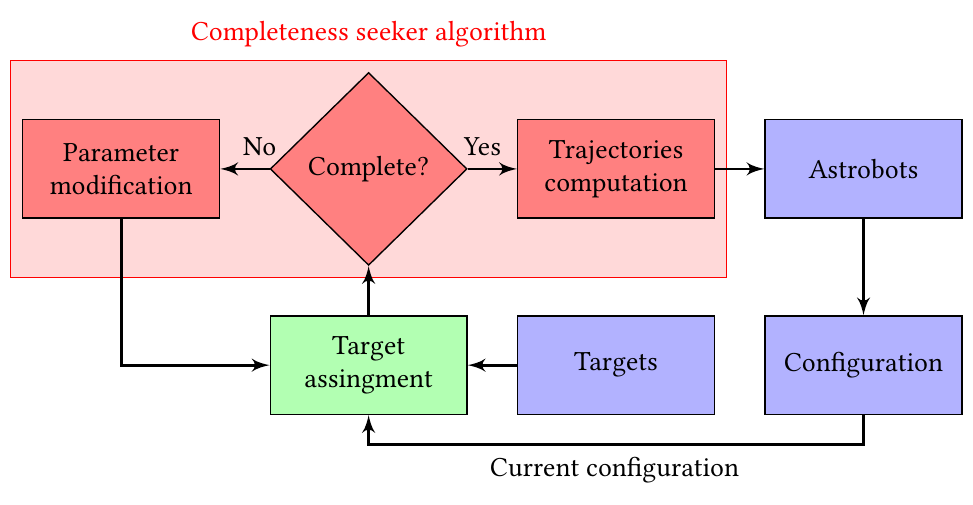}
	\caption{Multi-observation control strategy of astrobots. Targets may be selected from any astronomical catalog each of whose objects' projections can be covered by the patrol zone of at least one astrobot.}
	\label{fig:strategy}
\end{figure}

Thus, given the kinematic model in Eq. (\ref{eq:kin}) and CAPF in Eq. (\ref{eq:CAPF}), the control law associated with the $i$th astrobot reads as follows.
\begin{equation}
\label{eq:u}
\bm{u^{i}} := -\nabla_{\theta_{i},\phi_{i}}\xi(\bm{q^{i}})
\end{equation}	
A coordination process continues until the gradient in Eq. (\ref{eq:u}) vanishes implying that the corresponding $i$th astrobot has reached its target. \citet{macktoobian2019complete} showed that the global full convergence of a swarm of astrobots is achieved if the following completeness condition is fulfilled for every neighborhood of that swarm
\begin{equation}
\label{eq:cond}
\bm{q_\mathcal{T}} = -(\bm{\Lambda} + \bm{\Gamma})^{-1}\bm{\Theta},
\end{equation}  
where $\bm{q_\mathcal{T}}$ is the target position matrix of a neighborhood, and the remainder of the matrices are functions of various constituents of Eq. (\ref{eq:CAPF}). In particular, $\bm{\Lambda}$ is a function of attractive and cooperative attractive terms. This matrix represents the required force toward the elements of $\bm{q_\mathcal{T}}$. In contrast, $\bm{\Theta}$ and $\bm{\Gamma}$ denote the dynamic and static inhibitions to guarantee the safety of the interactions in the neighborhood. Given a particular setting of astrobots' configurations and target locations associated with an observation, one first has to check whether Eq. (\ref{eq:cond}) holds for every neighborhood of a desired focal plane. If the conditions set holds, then the CAPFs of astrobots generate the trajectories which yield to the convergence of all astrobots. Otherwise, the astrobots' initial configurations and/or their assignment to targets have to be revised for the purpose of fulfilling the completeness conditions for various neighborhoods. The explained procedure is schematically represented in Fig. \ref{fig:strategy}.
\subsection{Setup}
\label{subsec:setup}
\begin{figure}[!htp]
	\centering
	\includegraphics[scale=0.4]{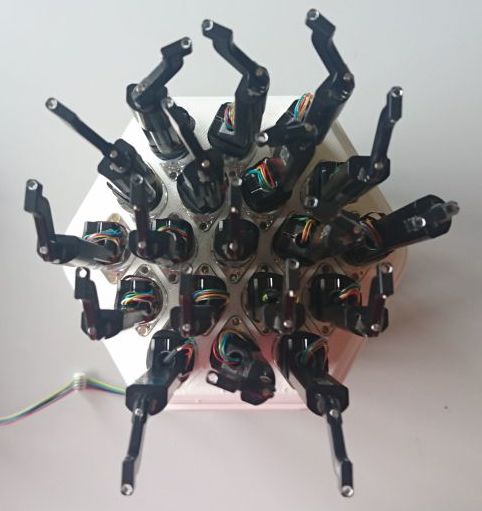}
	\caption{The 19-astrobot setup of the applied experimental tests. The characteristics of the 3D printed focal plane resemble the optical and the mechanical properties of the real focal planes of the SDSS telescopes such as pitch and focal distances of optical fibers.}
	\label{fig:swarm}
\end{figure}
Our hardware setup includes a miniature plate into which 19 astrobots are mounted (Fig. \ref{fig:swarm}). The relative distances between astrobots on this plate resemble those corresponding to the real focal planes of both SDSS-V telescopes. Parameters characterization of astrobots used in our experimental tests are described in Table \ref{tbl:1}. The specification of CAPF parameters may differ from one coordination scenario to another because the completeness condition requires different setting of parameters to be fulfilled. Thus, we elaborate on various settings of parameters and their impacts on the convergence of the swarm in Section \ref{sec:res}. The trajectories are generated by CSA, running on an instrument control system (ICS). The result is a YAML file representing two arrays of velocity data per astrobot each of which is associated with the joint of one of the arms. This file is transferred from the ICS to a communication hub via an Ethernet cable. The communication hub sends the trajectories to a bridge from which a CAN cable feeds the trajectories to astrobots.

\begin{figure}[!htp]
	\centering
	\includegraphics[scale=0.85]{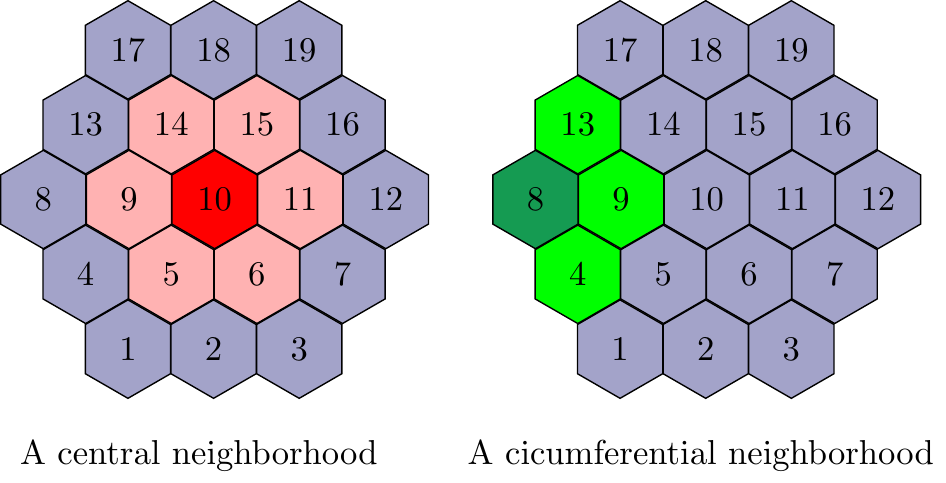}
	\caption{The neighborhood types in the experimental focal plane. In a massive focal plane, the majority of neighborhoods are central ones. Thus, the completeness condition, i.e., Eq. (\ref{eq:cond}), was obtained based on this critical case in \citep{macktoobian2019complete}.}
	\label{fig:neigh}
\end{figure}

In Section \ref{subsec:theo}, we noted that the completeness checking has be to done in a localized manner. The cardinality of astrobots in a neighborhood plays an important role in this analysis. In particular, full neighborhoods, which include 7 astrobots, are more prone to partial convergences compared to incomplete neighborhoods, i.e., those which are formed by less astrobots. So, we planned a radial placement of 19 astrobots as depicted in Fig. \ref{fig:neigh}. Each neighborhood is identified by its central astrobot. For example, the left-hand-side (full) neighborhood of Fig. \ref{fig:neigh} refers to that of astrobot \#10, while the (incomplete) right-hand-side one corresponds to astrobot \#8.

We conduct our tests according to a real galaxy catalog. Namely, we take various partitions of the eBOSS large-scale structure LRG catalog (data release 14) \citep{bautista2018sdss}\footnote{The data model can be found at \url{https://www.sdss.org/dr14/data_access/value-added-catalogs/?vac_id=eboss-large-scale-structure-lrg-catalogs-dr14}}. In the experimental tests, we randomly select \z{19} targets, from the data entries of the catalog, whose projections are in the area of the test plane. In each coordination The targets are assigned to astrobots using the optimal target assignment method \citep{macktoobian2020optimal}. So, we obtain the best possible performance in terms of the minimum effort for coordination and the maximum distribution of targets among astrobots to minimize the potential deadlock/collision situations. \citet{macktoobian2019complete} reported a theoretical framework to completely coordinate astrobots swarms whose cardinality are comparable to those of the SDSS-V telescopes and beyond. In this paper, we supply experimental and simulation tests to both SDSS-V and larger focal planes to validate the efficiency of CSA in safe completeness seeking in massive swarms of astrobots.   
\section{Results\protect\footnote{In the experiments of this paper, CSA is run on a Linux Manjaro 64-bit workstation with an Intel Core-i7 860 CPU, 16GB RAM, and an NVIDIA G94GL FX 1800 graphics card.}}
\label{sec:res}
\subsection{Parameters impacts on complete coordination}
\begin{figure}[b]
	\centering
	\includegraphics[scale=0.48]{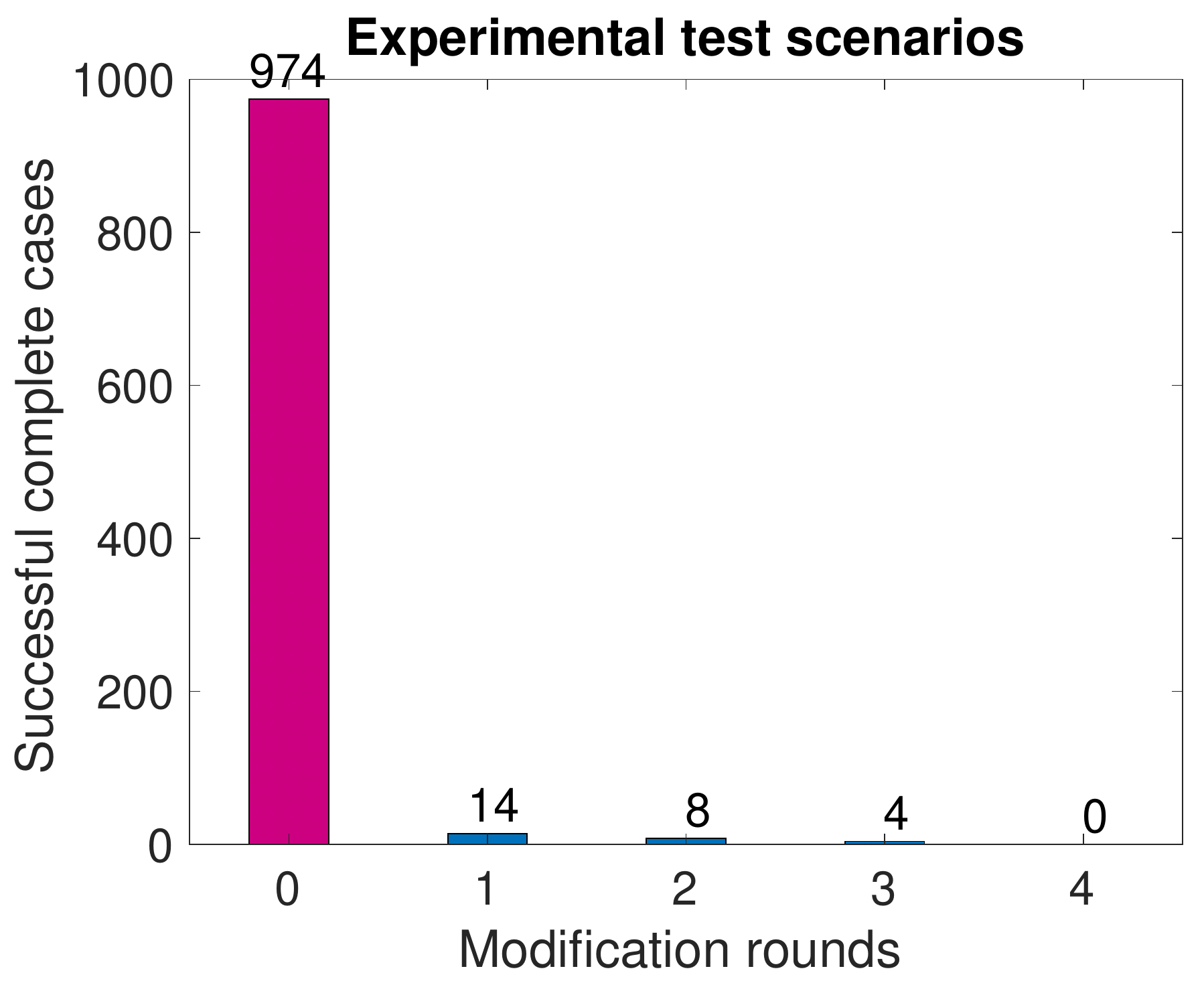}
	\caption[The number of the required modification rounds to reach completeness in 1000 experimental tests]{The number of the required modification rounds to reach completeness in 1000 experimental tests. The first bar indicates that 974 out of 1000 tested scenarios reached their full convergence without necessity of any parameter modification.}
	\label{fig:dead17}
\end{figure}

In this section\footnote{The coordination results of our experimental tests may be viewed at \url{http://y2u.be/MpXWvpz4h00.}}, we report the robustness of CSA in achieving completeness. If a particular setting of parameters do not end up with completeness, the theory requires the modification of those parameters to yield completeness. The intrinsic robustness of CSA in achieving completeness is investigated through 1000 tests applied to the 19-astrobot bench. Accordingly, Fig. \ref{fig:dead17} indicates that CSA reaches completeness in \%97.4 of the applied tests without any parameter modification. In other words, the cooperative kernels of the CAPFs used in CSA are efficiently sufficient to coordinate all of the astrobots given the initial configuration of the system and their targets. The results on the robustness notion may be extended to larger focal planes. Namely, Table \ref{tbl:dead} illustrates the simulated robustness results corresponding a class of massive focal planes. In this regard, one observes that even in the case of these extremely complicated focal planes, $\sim$\%97 of coordination scenarios are inherently complete. The completeness condition, i.e., Eq. (\ref{eq:cond}), is derived based on the local linearization of astrobot's motions in neighborhoods. Thus, the condition indeed approximates the completeness in a particular neighborhood. The quoted results echoes the efficiency of CSA in view of robustness in this viewpoint, as well. In particular, one observes that the applied linear approximations do not severely undermine the coordination quality in terms of the required modification rounds. To specifically study the impact of parameter variation in completeness seeking, we first note that the parameter $\bm{\Lambda}$ is exclusively a function of attractive weight $\lambda_{1}$ and cooperative attractive term $\lambda_{3}$. On the other hand, $\bm{\Gamma}$ and $\bm{\Theta}$ are extremely non-linear parameters which also include target positions. So, varying $\bm{\Lambda}$ may be preferred to $\bm{\Gamma}$ and $\bm{\Theta}$ because of its more intuitive definition. $\bm{\Lambda}$ modification can influence both safety and performance measures of a swarm. First, both constituents of $\bm{\Lambda}$ are attractive weights. So, increasing both of them may escalate the collision hazards because of the potential faster motions of astrobots' arms. Additionally, convergence time may inefficiently increase if one reduce both of these weights because the whole attractive dynamics of the swarm diminish. Thus, given a variation step $\delta \lambda >0$, we modify $\bm{\Lambda}$ based on the following rules.
\begin{equation}
\begin{split}
\lambda_{1} \leftarrow \lambda_{1} -\delta \lambda\\
\lambda_{3} \leftarrow \lambda_{3} +\delta \lambda\\
\end{split}
\end{equation}
This update profile preserves safety by decreasing the dominant weight factor, i.e., $\lambda_{1}$. The loss in the performance is also relatively compensated by increasing the submissive factor, say, $\lambda_{3}$.
\begin{table}[t]
	\centering
	\caption{Parameters of astrobots in the performed tests (The temporal step size can be manipulated as a degree of freedom in command generation. Smaller step sizes may provide more smooth and accurate motions specially in the maneuvers in which astrobots are so close to each other. On the other hand, very small step sizes increase the size of the command file by many redundant entries. So, this trade-off has to be managed by trial and error according to the cardinality of swarms of astrobots and their pitch.)}
	\begin{tabular}{l@{\hskip 0.4in}l@{\hskip 0.1in}}
		\toprule[2pt]\midrule  
		Parameters&Values\\\cmidrule(lr){1-1}\cmidrule{2-2}
		pack cardinality&\num{19}\\
		length of first arm&\SI{7.4}{\mm}\\
		length of second arm&\SI{15}{\mm}\\
		rotational step size&\ang{0.1}\\
		temporal step size&\SI{0.25}{\second}\\
		pitch & \SI{22.4}{\mm}\\
		\bottomrule[2pt]	
	\end{tabular}
	\label{tbl:1}
\end{table}
The smaller the update step $\delta\lambda$ is, the less safety is put in jeopardy. So, we report how effective the small variations of $\delta\lambda$ are in resolving incomplete coordination scenarios. According to Fig. \ref{fig:dead17}, only 26 scenarios were not inherently coordinated with respect to initial system's configuration. To resolve these issues, we define an step array $\overline{\delta\lambda}$ whose entries represent potential updates steps to be used.
\begin{equation}
\overline{\delta\lambda} \coloneqq \{0.001, 0.005, 0.01, 0.05, 0.1\}
\end{equation}
\begin{table}[t]
	\renewcommand{\arraystretch}{1.5}
	\setlength{\tabcolsep}{3.5mm}
	\centering
	\caption{The number of the required modification rounds to reach completeness in 1000 simulated tests (The first column of the modification rounds refers to the coordination which are complete, thereby needing no modifications. The large numbers of this column relative to the other columns per row exhibit the high performance of CSA in achieving completeness without modifications in the majority of situations.)}
	\begin{tabular}{ccccccc}
		\toprule[2pt]\midrule  
		\multirow{2}{*}{Population}&\multicolumn{6}{c}{Modification rounds}\\\cmidrule(lr){2-7}&0&1&2&3&4&5\\\cmidrule(lr){1-1}\cmidrule(lr){2-2}\cmidrule(lr){3-3} \cmidrule(lr){4-4}\cmidrule(lr){5-5}\cmidrule(lr){6-6}\cmidrule(lr){7-7}
		500&490&8&2&0&0&0\\
		1000&973&19&6&2&0&0\\
		3000&2910&66&15&8&1&0\\
		5000&4803&122&64&14&5&2\\
		\bottomrule[2pt]		
	\end{tabular}
	\label{tbl:dead}
\end{table}
\begin{figure}[b]
	\centering
	\includegraphics[scale=0.48]{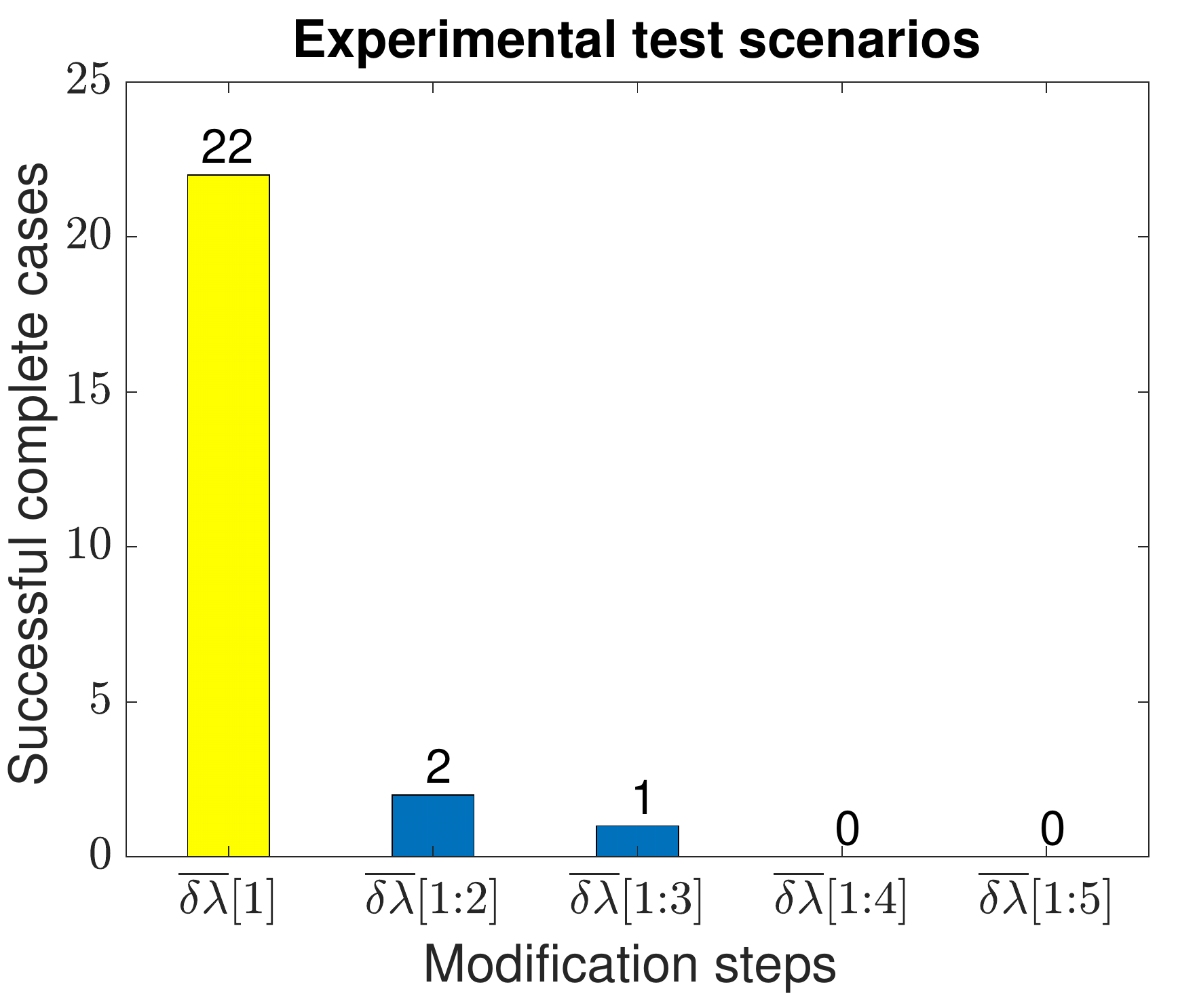}
	\caption[The number of the required modification steps to reach completeness in 1000 experimental tests]{The number of the required modification steps to reach completeness in 1000 experimental tests. The symbol $\overline{\delta\lambda}[1:n]$ denotes that the entries 1 to $n-1$ of the step array $\overline{\delta\lambda}$ could not resolve an incompleteness but the $n$th entry does it. The first bar of this figure refers to $\overline{\delta\lambda}[1]$ stating that the majority of incomplete cases were simply resolved using the first entry of $\overline{\delta\lambda}$.}
	\label{fig:lambda}
\end{figure}
Since we are interested in the smallest possible variation which fixes an incompleteness, the array is sorted in ascending order. Then, we pick the entries and feed them into a problematic completeness condition. If that new parameter does not resolve the intended incompleteness, the next one is picked to be tested, and so on, to finally find a new parameter to meet the completeness condition. In Fig. \ref{fig:lambda}, we observe that 22 out of 25 incomplete cases of the 19-astrobot bench are resolved using the smallest entry of the step array, say, $\overline{\delta\lambda} [1]= 0.001$. The second bar of Fig. \ref{fig:lambda} indicates 2 cases which were handled not by the first but by the second entry of the array. So, two modifications step have to be taken into account to first check $\overline{\delta\lambda} [1]= 0.001$ and then $\overline{\delta\lambda} [2]= 0.005$. Overall, there existed only one case in which the first two steps were not able to provide completeness but the third one. No incomplete scenario required larger steps, say, the two last entries of the array. 
\subsection{Convergence time}
Coordination are in general conducted from one observation to another. For this purpose, there are two approaches to reconfigure astrobots. The first strategy directly coordinates them from the configuration of the latest observation to that of the upcoming one. This direct convergence is generally fast, but it may be fairly challenging in terms of collision avoidance. Another scheme is a two-phase coordination in which astrobots are first sent to their fully folded state in which $\theta = \ang{0}$ and $\phi = \ang{\pi}$. Then, they are coordinated to their target configuration. The advantage of this strategy is that astrobots may encounter less potential deadlock situations. However, this way of coordination is at the cost of longer times to reach final coordination. This idea also implies more fluctuations thereby requiring more energy. So, in long runs, astrobots may be more prone to amortization.

\begin{figure}
	\centering
	\includegraphics[scale=0.48]{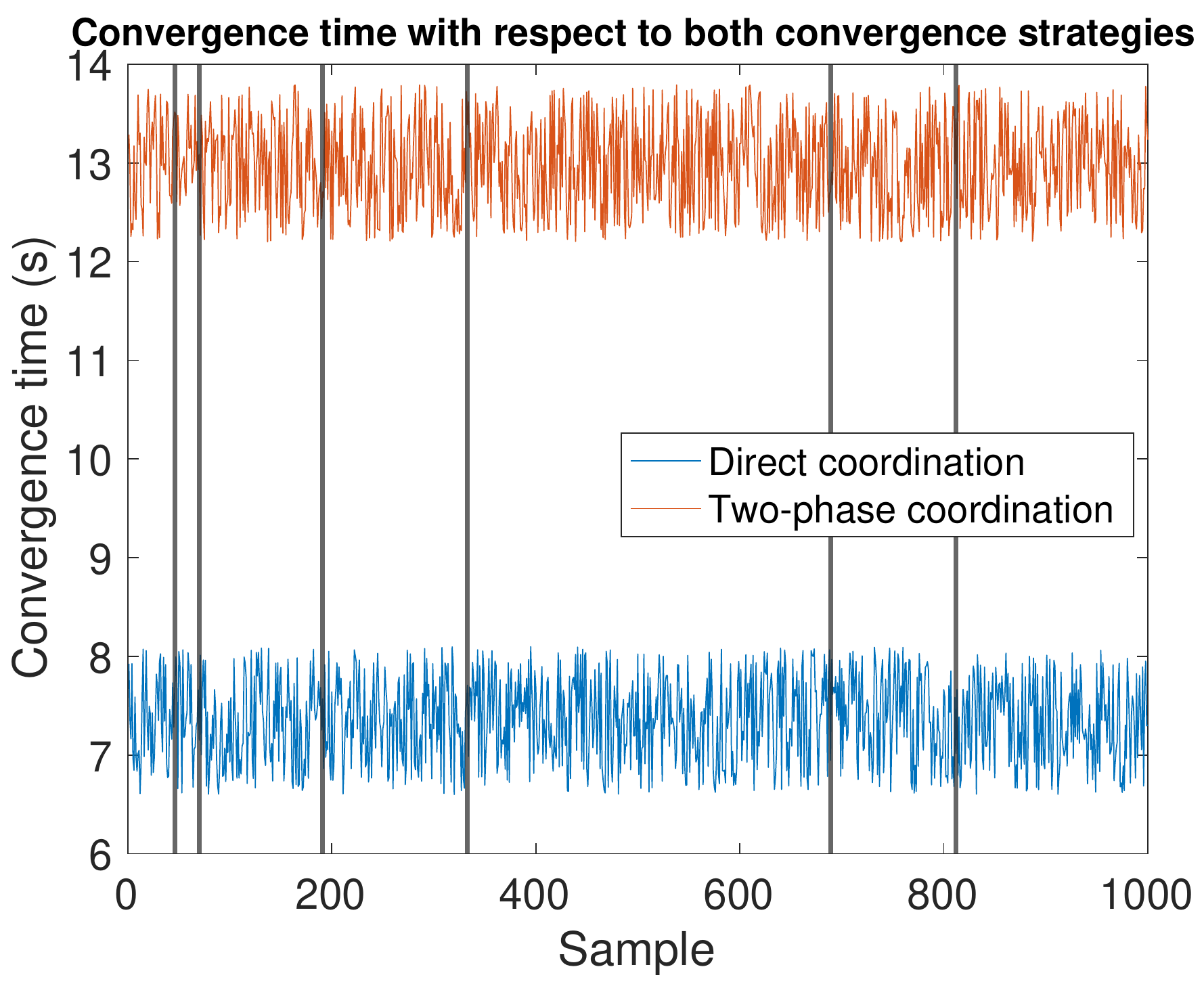}
	\caption[Convergence times of 1000 tests on the experimental setup with respect to both coordination strategies]{Convergence times of 1000 tests on the experimental setup with respect to both coordination strategies. The vertical black lines represent those scenarios which were complete in two-phase coordination. However, they were complicated enough to require one round of parameter modification to be also complete in direct coordination. Given the minority of these cases (6 occurrences) compared to the overall number of the tests, the direct coordination is the favorite approach.}
	\label{fig:time}
\end{figure}

The convergence times of both methods associated with 1000 coordinated scenarios on the 19-astrobot bench are represented in Fig. \ref{fig:time}. One notes that the direct coordination under the control of CSA are noticeably faster than those executed in the two-phase way. The completeness difference between the two is trivial, in that the two-phase strategy had only achieved six complete coordination more than the direct one. However, all these cases were compensated by only one modification round of parameters. The corresponding samples are signified using dark vertical lines in Fig. \ref{fig:time}. So, CSA is efficient enough to simultaneously perform direct coordination and achieve high rates of completeness. 
\subsection{Target distribution influence on completeness}
In previous sections, we used optimal target assignment \citep{macktoobian2020optimal} which supplies the maximum distribution of astrobots, i.e., their safety, and the minimum coordination, i.e., the minimum effort and time, required to arrange them in a desired coordination. In this section, we illustrate that quality of the coordination results of CSA are even resilient to various target distributions. In the sections above, whenever we wanted to supply some targets to coordination computations, we picked a random subset of the targets available in the cluster galaxy catalog, presented by \citet{bautista2018sdss}, used in this study. The only selection condition was the reachability of each of those targets by at least one of the astrobots of our tests. In this section, however, we select targets in various unbalanced scenarios in some of which the targets may be distributed in non-uniform fashion over our focal plane. Intuitively, if targets are uniformly dispersed, it is more likely that each astrobot reaches more than one target. Thus, optimal target assignment may enjoy more flexibility in terms of matching astrobots to targets. In contrast, a more biased distribution of targets may degrade the quality of the optimal target assignment, thereby negatively impact the coordination phase. Such biased distributions increase the density of targets in various spots of a focal plane. So, the question is whether CSA may have difficulties to deal with the coordination of astrobots in such dense localities. The following results indeed investigate this question showing that the sensitivity of the CSA performance to the target distribution is not noticeable.

We first define a uniformly distributed set of targets in polar coordinate system $(r,\theta)$ whose center is assumed to be located at the base of the central astrobot of a swarm, which is astrobot \#8 in our experimental test bench.
\begin{figure}[t]
	\centering
	\includegraphics[scale=0.48]{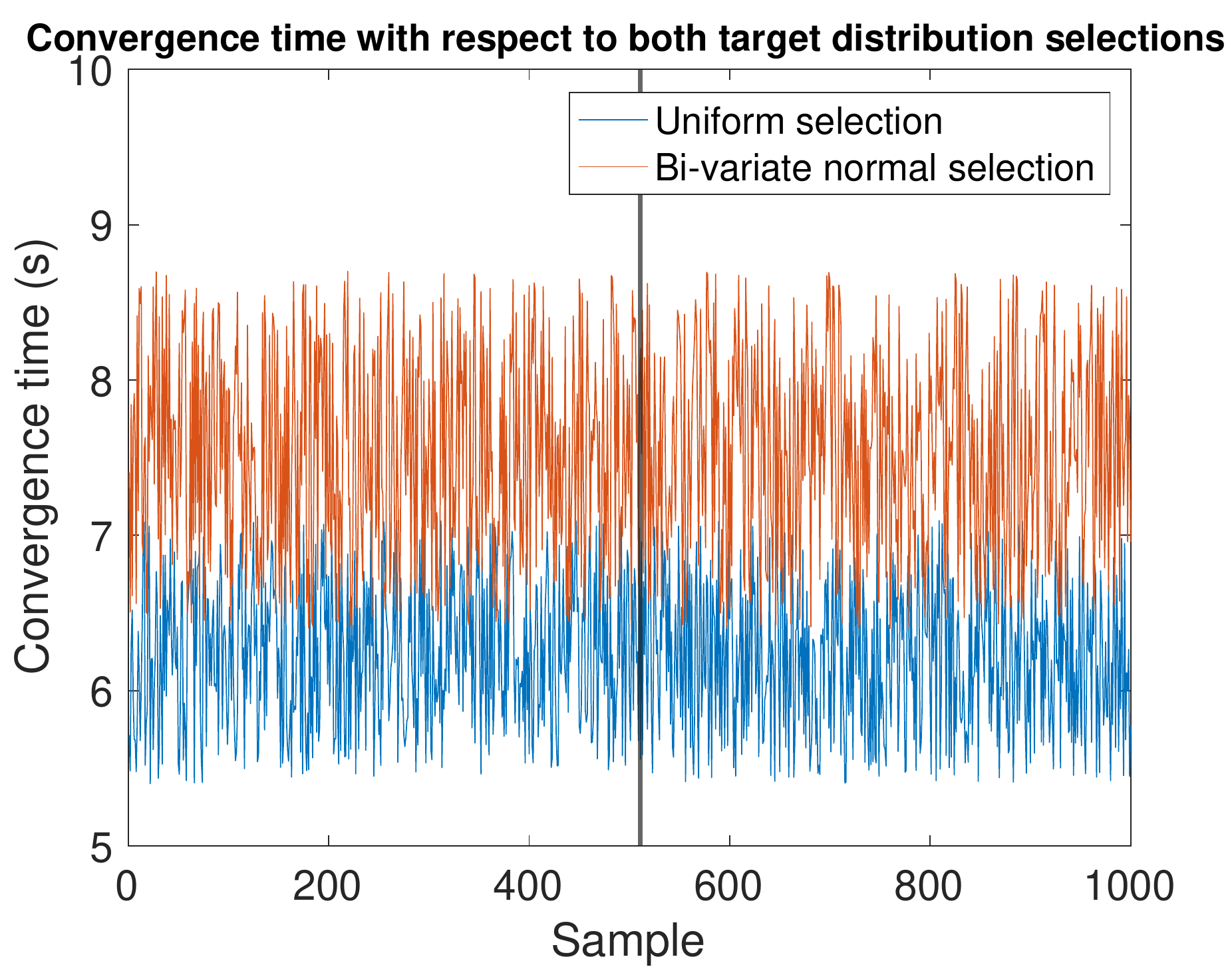}
	\caption[Convergence times of 1000 tests on the experimental setup with respect to both distribution selections]{Convergence times of 1000 tests on the experimental setup with respect to both distribution selections. Uniform selections of targets are simpler to be handled by the optimal target assignment because of the availability of the more pairing options between astrobots and targets. The faster convergences of this selection class are shown in the figure. Nevertheless, the applied bi-variate normal selections' convergence times are noticeably comparable with those of the uniform selections. Given the fact that target assignments and coordination under bi-variate normal selections are more challenging, we conclude that CSA is efficient under various potential distributions of targets.}
	\label{fig:timedist}
\end{figure}
\begin{equation}
\begin{split}
&\theta \sim \mathcal{U}[-\pi;\pi]\\
&r^2 \sim \mathcal{U}(0;r^{2}_{\mathrm{max}})
\end{split}
\end{equation}
In the equations above, $\mathcal{U}$ denotes a uniform distribution generator, and $r_{\mathrm{max}}$ represents the radius of the focal plane which reads \SI{44.8}{\mm} in the case of our test bench.

We also take a bi-variate normal distribution into account such that the maximum concentration of targets are around the center of the swarm, and the distribution radially degrades while one moves toward the edge of the focal plane in any direction. The probability density function of this bi-variate normal distribution, in Cartesian coordinate system, is defined as
\begin{equation}
N(x,y)\sim\frac{1}{2 \pi \sqrt{1-\rho^2}} \exp \Big\{-\frac{1}{2 (1-\rho^2)} \big[ x^2-2\rho x y+y^2 \big] \Big\},
\end{equation}
in which variances are assumed to be 1, and the correlation coefficient $\rho$ equals 0.7.

These two distributions are applied to the galaxy catalog of targets as masks to filter those targets which are placed in the patterns similar to the intended distributions. After performing 1000 direct coordination scenarios for each of the uniform and bi-variate normal distributions, the convergence time results are obtained as depicted in Fig. \ref{fig:timedist}. In particular, the average coordination time corresponding to bi-variate normal selection is slightly longer than that of the coordination associated with the uniform selection. However, the performance of CSA in the more crucial case (bi-variate normal selection) closely follows the coordination speed of the less-problematic one (uniform selection). Furthermore, there was only one coordination scenario which was not complete even after one round of parameter modification in the crucial case, whereas it was fully achieved in the uniform case. Thus, we conclude that CSA is efficiently capable of yielding fast and safe coordination even in the case of biased selections of targets in view of their spatial positions.
\subsection{Discussion}
\z{The results presented in this paper are all based on the assumption that all of the astrobots of a focal plane are functional, thereby participating in their coordination process. However, in a realistic scenario, some astrobots may be disabled for performance or reliability issues. In such cases, the complexity of a coordination process extremely increases because each disabled astrobot may invariably block a considerable percentages of the working spaces of its neighboring peers. So, the set of possible safe trajectories corresponding to such a scenario may be severely constrained. In the case of partial functionality of a pack of swarms, the naive usage of CSA generally leads to extremely inadequate convergence rates. To alleviate this issue, one may imagine two workarounds. First, disabled astrobots have to be removed from the focal plane. So, they do not occupy any working space associated with their adjacent astrobots. Thus, coordination processes need no extra considerations. The drawback of this strategy is that the removal of astrobots, especially in massive swarms, may be time-consuming and labor-intensive.}

\z{Alternatively, one may add another constraint to target assignment process, say, given a functional astrobot next to a disabled one, no target located on the opposite side of the disabled astrobot, with respect to the functional one, shall be assigned to the functional astrobot. This supplementary constraint may relax the necessity of accessing the blocked space behind the disabled astrobot by the functional one. So, The trajectory search space of CSA may become smaller, thus achieving faster convergence of the whole swarm. However, the consideration of the quoted requirement in crowded neighborhoods around disabled astrobots may give rise to the appearance of orphan targets, i.e., those targets which cannot be assigned to any astrobot.}      
\section{Conclusion}
\label{sec:conc}
\z{Throughput} maximization in massive spectroscopic observations yields the highest possible resolutions for resulting surveys. The current generation of these surveys are being generated using hundreds to thousands of fibers each of which has to point a new target from one observation to another. Astrobots have been used to automatically coordinate fibers in short amounts of time in a safe manner. However, given the dense structure of astrobots in focal planes, collision avoidance operations have been emerged as barriers in reaching complete convergence of astrobots in previous spectroscopic surveys. In this paper, we took the idea of cooperative artificial potential fields to generally resolve incompleteness coordination. In particular, we examined this strategy as a completeness seeker algorithm using a pack of astrobots similar to the focal planes of the SDSS-V telescopes.

The experimental results illustrated how the algorithm is efficient in achieving completeness in various settings of astrobots configurations and target assignments. We assessed the impact of parameter variations in resolving time-to-time incomplete scenarios by the minimum number of iterations and potential hazards to the safety of astrobots swarms. In view of convergence time, we observed that our algorithm can practically manage the safety through fast direct coordination. The robustness of our strategy considering various distributions of targets was also investigated. In particular, biased distributions of targets whose coordinates are not uniform often make a coordination critical for traditional planners. However, we validated the successful functionality of the proposed method in two uniform and bi-variate normal distributions.
\begin{acknowledgements}
This work was financially supported by the Swiss National Science Foundation (SNF) grant number 20FL21\_185771 and the SLOAN ARC/EPFL agreement number SSP523. \z{The authors also thank the anonymous reviewer who provided helpful comments on the earlier draft of this paper.}
\end{acknowledgements}
\section*{Conflict of interest}
The authors declare that they have no conflict of interest.

\bibliographystyle{spbasic}      % basic style, author-year citations

\bibliography{references}

\begin{thebibliography}{40}
\providecommand{\natexlab}[1]{#1}
\providecommand{\url}[1]{{#1}}
\providecommand{\urlprefix}{URL }
\expandafter\ifx\csname urlstyle\endcsname\relax
  \providecommand{\doi}[1]{DOI~\discretionary{}{}{}#1}\else
  \providecommand{\doi}{DOI~\discretionary{}{}{}\begingroup
  \urlstyle{rm}\Url}\fi
\providecommand{\eprint}[2][]{\url{#2}}

\bibitem[{Alam et~al.(2015)Alam, Albareti, Prieto, Anders, Anderson, Anderton,
  Andrews, Armengaud, Aubourg, Bailey et~al.}]{alam2015eleventh}
Alam S, Albareti FD, Prieto CA, Anders F, Anderson SF, Anderton T, Andrews BH,
  Armengaud E, Aubourg {\'E}, Bailey S, et~al. (2015) The eleventh and twelfth
  data releases of the sloan digital sky survey: final data from sdss-iii. The
  Astrophysical Journal Supplement Series 219(1):12

\bibitem[{Bautista et~al.(2018)Bautista, Vargas-Maga{\~n}a, Dawson, Percival,
  Brinkmann, Brownstein, Camacho, Comparat, Gil-Mar{\'\i}n, Mueller
  et~al.}]{bautista2018sdss}
Bautista JE, Vargas-Maga{\~n}a M, Dawson KS, Percival WJ, Brinkmann J,
  Brownstein J, Camacho B, Comparat J, Gil-Mar{\'\i}n H, Mueller EM, et~al.
  (2018) The sdss-iv extended baryon oscillation spectroscopic survey: baryon
  acoustic oscillations at redshift of 0.72 with the dr14 luminous red galaxy
  sample. The Astrophysical Journal 863(1):110

\bibitem[{Blake et~al.(2011)Blake, Brough, Colless, Contreras, Couch, Croom,
  Davis, Drinkwater, Forster, Gilbank et~al.}]{blake2011wigglez}
Blake C, Brough S, Colless M, Contreras C, Couch W, Croom S, Davis T,
  Drinkwater MJ, Forster K, Gilbank D, et~al. (2011) The wigglez dark energy
  survey: the growth rate of cosmic structure since redshift z= 0.9. Monthly
  Notices of the Royal Astronomical Society 415(3):2876--2891

\bibitem[{Bundy et~al.(2014)Bundy, Bershady, Law, Yan, Drory, MacDonald, Wake,
  Cherinka, S{\'a}nchez-Gallego, Weijmans et~al.}]{bundy2014overview}
Bundy K, Bershady MA, Law DR, Yan R, Drory N, MacDonald N, Wake DA, Cherinka B,
  S{\'a}nchez-Gallego JR, Weijmans AM, et~al. (2014) Overview of the sdss-iv
  manga survey: mapping nearby galaxies at apache point observatory. The
  Astrophysical Journal 798(1):7

\bibitem[{Cirasuolo et~al.(2014)Cirasuolo, Afonso, Carollo, Flores, Maiolino,
  Oliva, Paltani, Vanzi, Evans, Abreu et~al.}]{cirasuolo2014moons}
Cirasuolo M, Afonso J, Carollo M, Flores H, Maiolino R, Oliva E, Paltani S,
  Vanzi L, Evans C, Abreu M, et~al. (2014) Moons: the multi-object optical and
  near-infrared spectrograph for the vlt. In: Ground-based and airborne
  instrumentation for astronomy V, International Society for Optics and
  Photonics, vol 9147, p 91470N

\bibitem[{Cirasuolo et~al.(2016)Cirasuolo, Consortium
  et~al.}]{cirasuolo2016moons}
Cirasuolo M, Consortium M, et~al. (2016) Moons: A new powerful multi-object
  spectrograph for the vlt. ASPC 507:109

\bibitem[{Cui et~al.(2012)Cui, Zhao, Chu, Li, Li, Zhang, Su, Yao, Wang, Xing
  et~al.}]{cui2012large}
Cui XQ, Zhao YH, Chu YQ, Li GP, Li Q, Zhang LP, Su HJ, Yao ZQ, Wang YN, Xing
  XZ, et~al. (2012) The large sky area multi-object fiber spectroscopic
  telescope (lamost). Research in Astronomy and Astrophysics 12(9):1197

\bibitem[{Das and D{\"u}lger(2005)}]{das2005mathematical}
Das MT, D{\"u}lger LC (2005) Mathematical modelling, simulation and
  experimental verification of a scara robot. Simulation Modelling Practice and
  Theory 13(3):257--271

\bibitem[{Dawson et~al.(2012)Dawson, Schlegel, Ahn, Anderson, Aubourg, Bailey,
  Barkhouser, Bautista, Beifiori, Berlind et~al.}]{dawson2012baryon}
Dawson KS, Schlegel DJ, Ahn CP, Anderson SF, Aubourg {\'E}, Bailey S,
  Barkhouser RH, Bautista JE, Beifiori A, Berlind AA, et~al. (2012) The baryon
  oscillation spectroscopic survey of sdss-iii. The Astronomical Journal
  145(1):10

\bibitem[{Dawson et~al.(2016)Dawson, Kneib, Percival, Alam, Albareti, Anderson,
  Armengaud, Aubourg, Bailey, Bautista et~al.}]{dawson2016sdss}
Dawson KS, Kneib JP, Percival WJ, Alam S, Albareti FD, Anderson SF, Armengaud
  E, Aubourg {\'E}, Bailey S, Bautista JE, et~al. (2016) The sdss-iv extended
  baryon oscillation spectroscopic survey: Overview and early data. The
  Astronomical Journal 151(2):44

\bibitem[{Dey et~al.(2019)Dey, Schlegel, Lang, Blum, Burleigh, Fan, Findlay,
  Finkbeiner, Herrera, Juneau et~al.}]{dey2019overview}
Dey A, Schlegel DJ, Lang D, Blum R, Burleigh K, Fan X, Findlay JR, Finkbeiner
  D, Herrera D, Juneau S, et~al. (2019) Overview of the desi legacy imaging
  surveys. The Astronomical Journal 157(5):168

\bibitem[{Ferraro et~al.(2019)Ferraro, Wilson, Abidi, Alonso, Ansarinejad,
  Armstrong, Asorey, Avelino, Baccigalupi, Bandura
  et~al.}]{ferraro2019inflation}
Ferraro S, Wilson MJ, Abidi M, Alonso D, Ansarinejad B, Armstrong R, Asorey J,
  Avelino A, Baccigalupi C, Bandura K, et~al. (2019) Inflation and dark energy
  from spectroscopy at $ z> 2$. arXiv preprint arXiv:190309208

\bibitem[{Forero-Romero et~al.(2009)Forero-Romero, Hoffman, Gottl{\"o}ber,
  Klypin, and Yepes}]{forero2009dynamical}
Forero-Romero J, Hoffman Y, Gottl{\"o}ber S, Klypin A, Yepes G (2009) A
  dynamical classification of the cosmic web. Monthly Notices of the Royal
  Astronomical Society 396(3):1815--1824

\bibitem[{Gunn et~al.(2006)Gunn, Siegmund, Mannery, Owen, Hull, Leger, Carey,
  Knapp, York, Boroski et~al.}]{gunn20062}
Gunn JE, Siegmund WA, Mannery EJ, Owen RE, Hull CL, Leger RF, Carey LN, Knapp
  GR, York DG, Boroski WN, et~al. (2006) The 2.5 m telescope of the sloan
  digital sky survey. The Astronomical Journal 131(4):2332

\bibitem[{Hamuy et~al.(2006)Hamuy, Folatelli, Morrell, Phillips, Suntzeff,
  Persson, Roth, Gonzalez, Krzeminski, Contreras et~al.}]{hamuy2006carnegie}
Hamuy M, Folatelli G, Morrell NI, Phillips MM, Suntzeff NB, Persson S, Roth M,
  Gonzalez S, Krzeminski W, Contreras C, et~al. (2006) The carnegie supernova
  project: The low-redshift survey. Publications of the Astronomical Society of
  the Pacific 118(839):2

\bibitem[{Hill and Lesser(1986)}]{hill1986deployment}
Hill J, Lesser MP (1986) Deployment of the mx spectrometer. In: Instrumentation
  in astronomy VI, International Society for Optics and Photonics, vol 627, pp
  303--320

\bibitem[{H{\"o}rler et~al.(2016)H{\"o}rler, Kronig, Kneib, Bleuler, and
  Bouri}]{horler201624mm}
H{\"o}rler P, Kronig L, Kneib JP, Bleuler H, Bouri M (2016) A 24mm diameter
  fibre positioner for spectroscopic surveys. In: Advances in Optical and
  Mechanical Technologies for Telescopes and Instrumentation II, International
  Society for Optics and Photonics, vol 9912, p 99125K

\bibitem[{Ivezi{\'c} et~al.(2008)Ivezi{\'c}, Sesar, Juri{\'c}, Bond, Dalcanton,
  Rockosi, Yanny, Newberg, Beers, Prieto et~al.}]{ivezic2008milky}
Ivezi{\'c} {\v{Z}}, Sesar B, Juri{\'c} M, Bond N, Dalcanton J, Rockosi CM,
  Yanny B, Newberg HJ, Beers TC, Prieto CA, et~al. (2008) The milky way
  tomography with sdss. ii. stellar metallicity. The Astrophysical Journal
  684(1):287

\bibitem[{Joyce et~al.(2016)Joyce, Lombriser, and Schmidt}]{joyce2016dark}
Joyce A, Lombriser L, Schmidt F (2016) Dark energy versus modified gravity.
  Annual Review of Nuclear and Particle Science 66:95--122

\bibitem[{Kimura et~al.(2010)Kimura, Maihara, Iwamuro, Akiyama, Tamura, Dalton,
  Takato, Tait, Ohta, Eto et~al.}]{kimura2010fibre}
Kimura M, Maihara T, Iwamuro F, Akiyama M, Tamura N, Dalton GB, Takato N, Tait
  P, Ohta K, Eto S, et~al. (2010) Fibre multi-object spectrograph (fmos) for
  the subaru telescope. Publications of the Astronomical Society of Japan
  62(5):1135--1147

\bibitem[{Kollmeier et~al.(2017)Kollmeier, Zasowski, Rix, Johns, Anderson,
  Drory, Johnson, Pogge, Bird, Blanc et~al.}]{kollmeier2017sdss}
Kollmeier JA, Zasowski G, Rix HW, Johns M, Anderson SF, Drory N, Johnson JA,
  Pogge RW, Bird JC, Blanc GA, et~al. (2017) Sdss-v: pioneering panoptic
  spectroscopy. arXiv preprint arXiv:171103234

\bibitem[{Macktoobian et~al.(2019{\natexlab{a}})Macktoobian, Gillet, and
  Kneib}]{macktoobian2019complete}
Macktoobian M, Gillet D, Kneib JP (2019{\natexlab{a}}) Complete coordination of
  robotic fiber positioners for massive spectroscopic surveys. Journal of
  Astronomical Telescopes, Instruments, and Systems 5(4):045002

\bibitem[{Macktoobian et~al.(2019{\natexlab{b}})Macktoobian, Gillet, and
  Kneib}]{macktoobian2019navigation}
Macktoobian M, Gillet D, Kneib JP (2019{\natexlab{b}}) The navigation of
  robotic fiber positioners in sdss-v project: design and implementation. In:
  2019 15th Conference on Ph. D Research in Microelectronics and Electronics
  (PRIME), IEEE, pp 85--88

\bibitem[{Macktoobian et~al.(2020)Macktoobian, Gillet, and
  Kneib}]{macktoobian2020optimal}
Macktoobian M, Gillet D, Kneib JP (2020) Optimal target assignment for massive
  spectroscopic surveys. Astronomy and Computing 30:100364

\bibitem[{Makarem et~al.(2014)Makarem, Kneib, Gillet, Bleuler, Bouri, Jenni,
  Prada, and Sanchez}]{makarem2014collision}
Makarem L, Kneib JP, Gillet D, Bleuler H, Bouri M, Jenni L, Prada F, Sanchez J
  (2014) Collision avoidance in next-generation fiber positioner robotic
  systems for large survey spectrographs. Astronomy \& Astrophysics 566:A84

\bibitem[{Mathews et~al.(2006)Mathews, Lomholt, and
  Littlewood}]{mathews2006method}
Mathews MB, Lomholt PT, Littlewood WA (2006) Method and system for distributed
  navigation and automated guidance. US Patent 7,149,625

\bibitem[{Riess et~al.(1998)Riess, Filippenko, Challis, Clocchiatti, Diercks,
  Garnavich, Gilliland, Hogan, Jha, Kirshner et~al.}]{riess1998observational}
Riess AG, Filippenko AV, Challis P, Clocchiatti A, Diercks A, Garnavich PM,
  Gilliland RL, Hogan CJ, Jha S, Kirshner RP, et~al. (1998) Observational
  evidence from supernovae for an accelerating universe and a cosmological
  constant. The Astronomical Journal 116(3):1009

\bibitem[{Schlegel et~al.(2019{\natexlab{a}})Schlegel, Kollmeier, and
  Ferraro}]{schlegel2019megamapper}
Schlegel D, Kollmeier JA, Ferraro S (2019{\natexlab{a}}) The megamapper: a z> 2
  spectroscopic instrument for the study of inflation and dark energy. BAAS
  51(7):229

\bibitem[{Schlegel et~al.(2019{\natexlab{b}})Schlegel, Kollmeier, Aldering,
  Bailey, Baltay, Bebek, BenZvi, Besuner, Blanc, Bolton
  et~al.}]{schlegel2019astro2020}
Schlegel DJ, Kollmeier JA, Aldering G, Bailey S, Baltay C, Bebek C, BenZvi S,
  Besuner R, Blanc G, Bolton AS, et~al. (2019{\natexlab{b}}) Astro2020 apc
  white paper: The megamapper: az> 2 spectroscopic instrument for the study of
  inflation and dark energy. arXiv preprint arXiv:190711171

\bibitem[{Scoccimarro(2004)}]{scoccimarro2004redshift}
Scoccimarro R (2004) Redshift-space distortions, pairwise velocities, and
  nonlinearities. Physical Review D 70(8):083007

\bibitem[{Seo and Eisenstein(2003)}]{seo2003probing}
Seo HJ, Eisenstein DJ (2003) Probing dark energy with baryonic acoustic
  oscillations from future large galaxy redshift surveys. The Astrophysical
  Journal 598(2):720

\bibitem[{Smee et~al.(2013)Smee, Gunn, Uomoto, Roe, Schlegel, Rockosi, Carr,
  Leger, Dawson, Olmstead et~al.}]{smee2013multi}
Smee SA, Gunn JE, Uomoto A, Roe N, Schlegel D, Rockosi CM, Carr MA, Leger F,
  Dawson KS, Olmstead MD, et~al. (2013) The multi-object, fiber-fed
  spectrographs for the sloan digital sky survey and the baryon oscillation
  spectroscopic survey. The Astronomical Journal 146(2):32

\bibitem[{Takada et~al.(2006)Takada, Komatsu, and
  Futamase}]{takada2006cosmology}
Takada M, Komatsu E, Futamase T (2006) Cosmology with high-redshift galaxy
  survey: neutrino mass and inflation. Physical Review D 73(8):083520

\bibitem[{Takada et~al.(2014)Takada, Ellis, Chiba, Greene, Aihara, Arimoto,
  Bundy, Cohen, Dor{\'e}, Graves et~al.}]{takada2014extragalactic}
Takada M, Ellis RS, Chiba M, Greene JE, Aihara H, Arimoto N, Bundy K, Cohen J,
  Dor{\'e} O, Graves G, et~al. (2014) Extragalactic science, cosmology, and
  galactic archaeology with the subaru prime focus spectrograph. Publications
  of the Astronomical Society of Japan 66(1)

\bibitem[{Tao et~al.(2018)Tao, Makarem, Bouri, Kneib, and
  Gillet}]{tao2018priority}
Tao D, Makarem L, Bouri M, Kneib JP, Gillet D (2018) Priority coordination of
  fiber positioners in multi-objects spectrographs. In: Ground-based and
  Airborne Instrumentation for Astronomy VII, International Society for Optics
  and Photonics, vol 10702, p 107028K

\bibitem[{Wake et~al.(2017)Wake, Bundy, Diamond-Stanic, Yan, Blanton, Bershady,
  S{\'a}nchez-Gallego, Drory, Jones, Kauffmann et~al.}]{wake2017sdss}
Wake DA, Bundy K, Diamond-Stanic AM, Yan R, Blanton MR, Bershady MA,
  S{\'a}nchez-Gallego JR, Drory N, Jones A, Kauffmann G, et~al. (2017) The
  sdss-iv manga sample: design, optimization, and usage considerations. The
  Astronomical Journal 154(3):86

\bibitem[{Way et~al.(2005)Way, Quintana, Infante, Lambas, and
  Muriel}]{way2005redshifts}
Way M, Quintana H, Infante L, Lambas D, Muriel H (2005) Redshifts in the
  southern abell redshift survey clusters. i. the data. The Astronomical
  Journal 130(5):2012

\bibitem[{Zasowski et~al.(2017)Zasowski, Cohen, Chojnowski, Santana, Oelkers,
  Andrews, Beaton, Bender, Bird, Bovy et~al.}]{zasowski2017target}
Zasowski G, Cohen R, Chojnowski SD, Santana F, Oelkers R, Andrews B, Beaton R,
  Bender C, Bird J, Bovy J, et~al. (2017) Target selection for the sdss-iv
  apogee-2 survey. The Astronomical Journal 154(5):198

\bibitem[{Zehavi et~al.(2002)Zehavi, Blanton, Frieman, Weinberg, Mo, Strauss,
  Anderson, Annis, Bahcall, Bernardi et~al.}]{zehavi2002galaxy}
Zehavi I, Blanton MR, Frieman JA, Weinberg DH, Mo HJ, Strauss MA, Anderson SF,
  Annis J, Bahcall NA, Bernardi M, et~al. (2002) Galaxy clustering in early
  sloan digital sky survey redshift data. The Astrophysical Journal 571(1):172

\bibitem[{Zhao et~al.(2012)Zhao, Zhao, Chu, Jing, and Deng}]{zhao2012lamost}
Zhao G, Zhao YH, Chu YQ, Jing YP, Deng LC (2012) Lamost spectral survey—an
  overview. Research in Astronomy and Astrophysics 12(7):723

\end{thebibliography}

\end{document}